\documentclass[10pt,twocolumn,pra,aps]{revtex4-2}

\usepackage{graphicx}
\usepackage{amsmath}
\usepackage{amsfonts}
\usepackage{amssymb}
\usepackage[version=3]{mhchem}
\usepackage{siunitx}
\usepackage[colorlinks=true,linkcolor=blue,citecolor=blue,urlcolor=blue]{hyperref}

\begin{document}

\title{Hybrid approach to reconstruct nano\-scale grating dimensions using scattering and fluorescence with soft X-rays}

% Authors and Affiliations
\author{Leonhard M. Lohr}
\email{leonhard.lohr@ptb.de}
\author{Richard Ciesielski}
\author{Vinh-Binh Truong}
\author{Victor Soltwisch}
\affiliation{Physikalisch-Technische Bundesanstalt (PTB), Abbestraße 2-12, 10587 Berlin, Germany}

\begin{abstract}
\noindent Scattero\-metry is a tested method for measuring periodic semiconductor structures. Since the sizes of modern semiconductor structures have reached the nano\-scale regime, the challenge is to determine the shape of periodic nano\-structures with sub-nanometer accuracy. To increase the resolution of scattero\-metry, short-wavelength radiation like soft X-rays can be used. But, scattero\-metry with soft X-rays is an inverse problem whose solutions can be ambiguous and its sensitivity should be further increased to determine the shape of even more complicated periodic nano\-structures made up of different materials. To achieve unique solutions with smaller uncertainties, scattero\-metry can leverage the excitation of low-Z materials with soft X-rays. Additional information from soft X-ray fluorescence analysis in a hybrid measurement approach can mitigate the problem of ambiguous solutions from soft X-ray scattering and could further decrease uncertainty. In this work, the hybrid approach is utilized to perform a comparison of solutions from the inverse problem and determine the actual solution over ambiguous solutions.
\end{abstract}

\maketitle

\vspace{5em}

\section{Introduction}
Due to the development and use of extreme ultraviolet (EUV) lithography since the last decade, the dimensions of semiconductor structures on integrated circuits have shrunk from the micro\-scale to the nano\-scale which increases the demand and importance of process control~\cite{IRDSMet:2024}. Wafer test structures, like periodic nano\-gratings or buried nano\-structures have feature sizes in the nanometer regime. Techniques like critical dimension scanning electron microscopy (CD-SEM)~\cite{Joy:2003}, scattero\-metry~\cite{Kleinknecht:1978,Naqvi:1992,McNeil:1993,Raymond:1995}, 3D atomic force microscopy (AFM)~\cite{Binnig:1986,Dixson:2003}, critical dimension small angle X-ray scattering (CD-SAXS)~\cite{Ho:2007}, transmission electron microscopy (TEM)~\cite{Knoll:1932,Sheng:1980} and scanning transmission electron microscope (STEM)~\cite{Ardenne:1938,Crewe:1969} probe such structures and measure their dimensions with accuracy from nanometer to sub-angstrom accuracy~\cite{Orji:2018}. Also, combining energy dispersive X-ray spectroscopy EDS~\cite{Fitzgerald:1968,Gatti:1984} with another technique like STEM allows for elemental mapping~\cite{Tsuneta:2002}. The shrinking dimensions and feature sizes of such structures remain a challenge especially for wafer manufacturing control that makes use of in-line metrology~\cite{EmamiNaeini:2008}. In-line metrology for wafer tests requires fast and non-destructive techniques. scattero\-metry does not require destructive cross-sectioning and is fast due its non-scanning nature, making it compatible for in-line metrology. Its sensitivity in characterizing periodic nano\-structures is limited by the penetration depth, which depends on the wavelength, the incident angle of the radiation and the materials. The wavelength ranges from the near-infrared range to soft X-ray spectral range, depending on the setup. Optical critical dimension (OCD) metrology~\cite{Guillaume:1985} is a specialized form of scattero\-metry that typically uses radiation in the near-infrared, visible, ultraviolet (UV) and deep ultraviolet (DUV) spectral range. Like UV scattero\-metry~\cite{Wurm:2017}, this method can determine the critical dimension of periodic nano\-structures with nanometer accuracy~\cite{Allen:2007}. The accuracy of scattero\-metry techniques increases by using short-wavelength radiation like EUV radiation for EUV scattero\-metry~\cite{FernandezHerrero:2021} or soft X-rays for grazing-incidence small angle X-ray scattering (GISAXS)~\cite{Levine:1989,Wernecke:2014}, measuring all available diffraction orders~\cite{Lohr:2023}. scattero\-metry is a technique that utilizes a model of the periodic nano\-structure to simulate the diffraction efficiencies observed during the measurement. Further improvements to scattero\-metry can be performed by taking roughness effects from the nano\-structure into account~\cite{Gross:2012,FernandezHerrero:2019,FernandezHerrero:2022} and applying statistical approaches to the dimensional reconstruction~\cite{Henn:2013}. In the form of EUV scattero\-metry or soft X-ray scattering, scattero\-metry has the potential to become a candidate as reference technique~\cite{Bodermann:2014}. First prototypes of EUV scattero\-meters that make use of EUV lab sources have already been commissioned~\cite{Ku:2016,Bahrenberg:2020,Porter:2023}. However, the lack of stable EUV lab sources still remains a challenge for a wider adaptation into in-line metrology.

As more sophisticated models such as rigorous coupled wave analysis (RCWA)~\cite{Moharam:1981,Moharam:1995a,Moharam:1995b,Lalanne:1996,Li:1997} are developed, the standing wave field from the interaction of incoming and outgoing waves in the scattering process can be seen as the underlying principle of scattero\-metry. The standing wave field is determined by the shape and elemental composition of the periodic nano\-structure and is influenced by near-field effects. Models describing the standing wave field are crucial in understanding and using near-field effects in scattero\-metry. One way to rigorously calculate this field is to solve the Maxwell's equations with the finite element method (FEM)~\cite{Urbach:1991,Bao:1995,Elschner:2002}. In particular, the FEM is preferred over other, faster methods such as RCWA because the FEM allows for setting the numerical precision and modeling details such as roundings and inclines of a nano\-grating shape~\cite{Soltwisch:2017}. Unlike with scattering-type scanning near-field optical microscopy (s-SNOM)~\cite{Zenhausern:1995,Siebenkotten:2024} in the infrared spectral range, the standing wave field of soft X-rays cannot be directly measured because of the short wavelength. So far, its far field, which is equivalent to the fast Fourier transform applied to the calculated standing wave field, can be measured in the form of diffraction intensities. When measuring the diffraction efficiency, the phase information is lost. Thus, solving the inverse problem based on soft X-ray scattering measurements can yield ambiguous solutions of the shape of the periodic nano\-structure, also known as multimo\-dalities.

On the other hand, other techniques such as grazing incidence X-ray fluorescence (GIXRF)~\cite{Yoneda:1971,Boer:1991} with soft X-rays are based on the characteristic soft X-ray fluorescence (XRF) radiation emitted, which can be described by excitation of the material based on the standing wave field of incoming and outgoing waves~\cite{Reinhardt:2014}. Like scattero\-metry, GIXRF measurements can yield ambiguous solutions. Approaches using additional information by combining GIXRF and X-ray reflectivity (XRR)~\cite{Parratt:1954} for improving the result are currently under development~\cite{Ingerle:2014,Hoenicke:2019,Ingerle:2020,Melhem:2023}. In the combined approach, the overall sensitivity is increased or enhanced by combining individual sensitivity. Also under development is the combination of some of the different techniques OCD, SEM, CD-SAXS, XRR, GISAXS and AFM~\cite{Yan:2006,Vaid:2011,Silver:2011,Silver:2014,Henn:2015,Griesbach_Schuch:2017,Siaudinyte:2023}, the combination of scattero\-metry and ellipsometry~\cite{Hansen:2022}, and the combination of GIXRF-XRR, near-edge X-ray absorption fine structure (NEXAFS)~\cite{Stoehr:1992} and GISAXS~\cite{Murataj:2023}. These first hybrid approaches are still in contrast to available setups with built-in hybrid measurement techniques such as EDS-STEM~\cite{Williams:2002,Herzing:2008}, which are potentially applicable for in-line reference metrology.

Scattero\-metry can make use of the fact that GIXRF is suitable for characterizing periodic nano\-structures made of low-Z materials with soft X-rays~\cite{Soltwisch:2018,Hoenicke:2020,Andrle:2021,Andrle:2023}. The description of the standing wave field of both methods is identical but it is assumed that the phenomena observed, soft X-ray scattering and soft X-ray fluorescence, have different sensitivities. Soft X-ray fluorescence can probe specific areas of a sample, yielding information about their spatial distribution as confirmation of the mass distribution, while soft X-ray scattering can get more information about the sample structure, depending on the optical contrast. As the combination of these techniques can yield a better representation of the structure and mass distribution inside the periodic nano\-structure, these methods are suitable for being combined in hybrid metrology. Recent research at PTB's soft X-ray beamline at BESSY II synchrotron facility has shown that a dedicated scattering chamber with a silicon drift detector (SDD) for collecting fluorescence spectra and a charge-coupled device (CCD) for capturing diffraction efficiency, can be used to collect hybrid measurement data from a sample volume of a nano\-scale grating~\cite{Ciesielski:2023}.

In this work, the multimo\-dality problem of scattero\-metry is shown to be mitigated by soft X-ray fluorescence scattero\-metry (XFS) as a hybrid measurement approach, using a modified version of this setup. In the hybrid approach, the setup combines soft X-ray scattering and soft X-ray fluorescence analysis for the dimensional reconstruction of a nano\-scale 1D line grating of silicon nitride (\ce{Si3N4}) and silicon dioxide (\ce{SiO2}). This work compares optimization results of dimensional reconstructions based on combined regressions using weighted soft X-ray scattering and soft X-ray fluorescence data from an angular scan and an appropriate model of the nano\-scale grating. With this method, this work aims to classify multimo\-dalities and find a range of appropriate weights for the data in the combined regression that can be used for hybrid dimensional reconstructions with minimized residuals.

\section{Experimental and fundamental}

\subsection{nano\-scale grating sample}
\label{sec:sample}

\begin{figure}[ht]
\centering
  \includegraphics[width=\columnwidth]{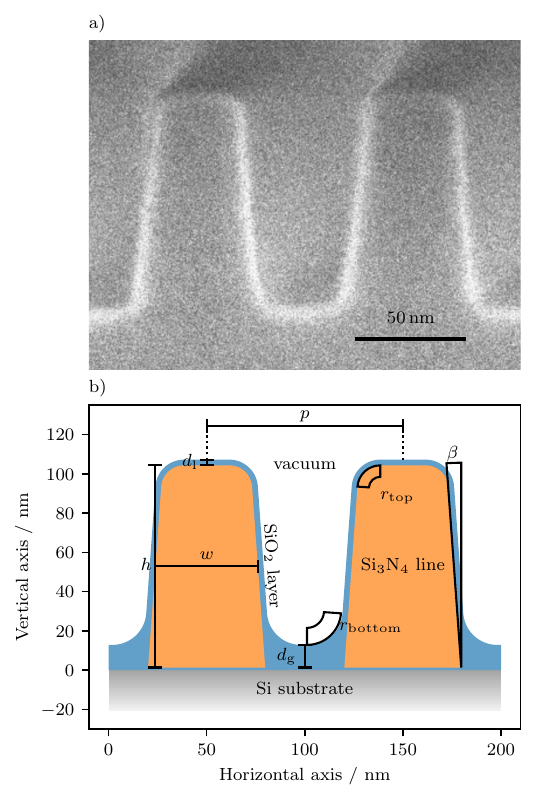}
  \caption{Cross-sectional view of the silicon nitride (\ce{Si3N4}) nano\-scale grating with a silicon dioxide (\ce{SiO2}) layer on a silicon (\ce{Si}) substrate. a) Scanning electron microscopy (SEM) image of the grating made at Helmholtz-Zen\-trum Berlin (HZB). b) The parametrization scheme of the shape of the grating profile from the SEM image.}
  \label{fgr:grating_profile}
\end{figure}

The hybrid approach of combining soft X-ray scattering and fluorescence makes use of the periodicity of nano\-structures and the fact that they are made of different materials with low-Z elements, like silicon dioxide, silicon nitride and carbon compounds, because their L- and K-edges are located in the soft X-ray spectral range. Thus, the sample to be characterized in this work is a 1D nano\-scale grating that consists of silicon nitride (\ce{Si3N4}) lines with a silicon dioxide (\ce{SiO2}) layer on a silicon (\ce{Si}) substrate. This grating was made at the Helmholtz-Zen\-trum Berlin (HZB) by means of electron beam lithography, applied to a \ce{Si} substrate with an \ce{Si3N4} layer on top. Figure~\ref{fgr:grating_profile}a shows an SEM image of the profile of this grating and Fig.~\ref{fgr:grating_profile}b the parametrization of the shape of the grating profile, identifying different materials that cannot be seen in Fig.~\ref{fgr:grating_profile}a. The pitch $p$ of this grating is $100\,\mathrm{nm}$, the nominal value for the critical dimension of the \ce{Si3N4} lines, here the width $w$ at half height, is $50\,\mathrm{nm}$ and the expected height, $h$, of the \ce{Si3N4} lines is $100\,\mathrm{nm}$. Other parameters are the sidewall angle $\beta$ of the \ce{Si3N4} lines, the substrate \ce{SiO2} layer thickness $d_\mathrm{substrate}$, the \ce{SiO2} layer thickness in the grooves of the lines $d_\mathrm{groove}$, the \ce{SiO2} layer thickness at the top of the \ce{Si3N4} lines $d_\mathrm{line}$, the bottom corner radius of the \ce{SiO2} layer $r_\mathrm{bottom}$, and the top corner radius of the \ce{Si3N4} lines $r_\mathrm{top}$. Here, the top corner radius of the \ce{SiO2} layer is determined by $r_\mathrm{top}+d_\mathrm{line}$ and, thus, dependent. The substrate \ce{SiO2} layer's thickness is known to be not equal to zero. Here, this thickness is fixed to $1.35\,\mathrm{nm}$. As the pitch is well-defined to be $p=100\,\mathrm{nm}$, the remaining set of varying parameters is $(h, w, \beta, r_\mathrm{bottom}, r_\mathrm{top}, d_\mathrm{groove}, d_\mathrm{line})$. A contamination layer of carbon compounds at the sample surface is well-known for significant absorption of fluorescence radiation from inner regions of the sample. Thus, before measuring the sample, the line grating sample in this work was cleaned at the University of Jena. The cleaning process started with Caro's acid, followed by an ultrasonic bath in an ammonia-water solution (1:150 \ce{NH3}:\ce{H2O}), water for rinsing, and spin dry. The cleaning process can cause minimal changes of the grating structure and can oxidize the surface.

\subsection{Theory of soft X-ray scattering on a nano\-scale gratings}
\label{sec:scat}

A nano\-scale grating, as shown in Fig.~\ref{fgr:grating_profile}, functions as a diffraction grating for soft X-rays. If monochro\-matized soft X-rays with the wavelength $\lambda$ hit the nano\-scale grating with the pitch $p$ under a grazing angle of incidence $\alpha_\mathrm{i}$ equal or smaller than the critical angle of total external reflection, almost all photons are elastically scattered into discrete diffraction maxima. For a grating in perfect conical mounting~\cite{Wernecke:2014}, the final angular positions of the diffraction maxima $\vartheta_\mathrm{f}$ and $\alpha_\mathrm{f}$ can be described as follows
\begin{align}
    \vartheta_\mathrm{f} &= \arctan\left(\frac{m\lambda / p}{\cos{\alpha_\mathrm{i}}}\right)\ ,\label{eq:thetaf}\\
    \alpha_\mathrm{f} &= \arcsin\left(\sqrt{\sin^2{\alpha_\mathrm{i}}-\left(m\lambda / p\right)^2}\right)\ ,\label{eq:alphaf}
\end{align}
where $m$ denotes the number of the diffraction order. These formulas are derived from the scattering vector $\mathbf{q} = \mathbf{k}_\mathrm{f} - \mathbf{k}_\mathrm{i}$ for perfect conical grating mounting, where $\mathbf{k}_\mathrm{i}$ is the wave vector of the incident photons and $\mathbf{k}_\mathrm{f}$ that of the scattered photons with $k_0 = |\mathbf{k}_\mathrm{i}| = |\mathbf{k}_\mathrm{f}| = 2\pi / \lambda$~\cite{Pflueger:2020}. When captured by an area detector, the diffraction pattern draws a half circle over the horizon of the grating plane (Fig.~\ref{fgr:hybrid_scheme}, left). The spacial displacements between the diffraction orders are determined by the grating pitch $p$, the grazing angle of incidence $\alpha_\mathrm{i}$ and the wavelength $\lambda$. The diffraction efficiency $I_m(\alpha_\mathrm{i}, E_\mathrm{i})$ over the diffraction order $m$ is determined by the shape of the nano\-scale grating profile and the optical properties of the grating materials. This measurand is proportional to the squared absolute value of the electric field strength far from the spot where scattering takes place. The standing wave field from incoming and outgoing waves at the grating, $\mathbf{E}(\mathbf{r}, y)$ with $\mathbf{r}=(x,z)^T$, shown in the coordinate system in Fig.~\ref{fgr:hybrid_scheme}, can be precisely calculated by solving Maxwell's equations with the FEM. Assuming a nano\-scale grating can be seen as infinitely long and has no defects in form of superstructure and varying pitch, a grating parametrization like shown in Fig.~\ref{fgr:grating_profile}b, the optical properties of the materials and the experimental parameters, $\alpha_\mathrm{i}$ and $\lambda$, can determine the electric field strength $\mathbf{E}(\mathbf{r}, y)$ of the standing wave field. Optical constants or the complex refractive index describe the optical properties of the grating materials for the interaction with soft X-rays.

Mathematically, the diffraction efficiency is proportional to the squared, Fourier-transformed electric field strength of the standing wave field $\left(I_m(\alpha_\mathrm{i}, E_\mathrm{i}) \propto |\mathbf{E}(k_{x,m},k_{z,m},y)|^2\right)$. An additional operation yields the diffraction efficiency as follows
\begin{equation}
    I_m(\alpha_\mathrm{i}, E_\mathrm{i}) \approx \left|\mathbf{E}(k_{x,m},k_{z,m},y)\right|^2 e^{-\xi^2 q_{x,m}^2 }\ ,
    \label{eq:diffraction_efficiency}
\end{equation}
where $\mathbf{E}(k_{x,m},k_{z,m},y)$ is normalized by the amplitude of the incoming plane waves. This definition contains an exponential damping factor, known as Debye-Waller damping factor, that is required to take the loss of intensity due to diffusely scattered photons at rough line edges into account~\cite{FernandezHerrero:2019}. This damping factor enhances the grating model with a kind of imperfection nano\-scale gratings typically have. Here, the roughness parameter $\xi$, measured in nm, indicates the variance of deviations of the edges and line widths of the grating lines. The horizontal component of the scattering vector $q_{x,m}$ of diffraction order $m$ determine the effect of the line edge and line width roughness to the diffraction efficiency.

\subsection{Theory of soft X-ray fluorescence from a nano\-grating}

Many characteristic fluorescence emission lines from low-Z materials lie in the soft X-ray spectral range. Therefore, GIXRF can be used for element-specific reconstruction of nano\-structures~\cite{Andrle:2021}. Depending on the incident photon energy $E_\mathrm{i}$ and the angular orientation of the grating with respect to the incident photon beam, elements of different materials of the grating can be excited by the incoming photons. Different parts of the grating then emit fluorescence photons whose spectrum contains characteristic lines of atoms excited. The intensity distribution of the electric field strength of the standing wave field $I(\mathbf{r}, y)\propto |\mathbf{E}(\mathbf{r},y)|^2$, calculated as explained in Sec.~\ref{sec:scat}, determines how strong different areas of the grating gross-section are stimulated. Thus, the fluorescence intensity depends on the excited area of the material $A_j$ in the cross-section of the grating. The fluorescence intensity of a certain emission line $l$ of an element within material $j$ can be described by the adapted and simplified Sherman equation~\cite{Sherman:1955}, according to:
\onecolumngrid
\begin{equation}
    \Phi_l(\alpha_\mathrm{i}) = \frac{w_i\rho_j\tau(E_\mathrm{i})\omega_k}{p}\iint_{A_j} |\mathbf{E}(\mathbf{r}, y)|^2 e^{-\sum_{j=1}^M\rho_j\mu_j(E_\mathrm{l})d_j(\mathbf{r})}\mathrm{d}\mathbf{r}\ ,
    \label{eq:sherman}
\end{equation}
\twocolumngrid
where $\mathbf{E}(\mathbf{r}, y)$ is normalized by the amplitude of the incoming plane waves. The mass densities of all $M$ materials, $\rho_j$ with $j\in [1, \dots, M]$, are assumed to be constant over their areas of the grating profile $A_j$ in an unit cell, whose width is given by the pitch $p$ of the grating. Fundamental parameters that determine how the integral scales to the absolute fluorescence intensity are the mass fraction of the fluorescence emitting element, denoted as $w_i$, the photo-ionization cross section of the relevant atomic shell for the incident photon energy $E_\mathrm{i}$, denoted as $\tau (E_\mathrm{i})$, the fluorescence yield of the relevant atomic shell, $\omega_k$, and the mass attenuation coefficient of material $j$ for the fluorescence energy of the line $l$, denoted as $\mu_j(E_l)$. The exponential term describes the self-absorption of fluorescence photons of emission line $l$ when passing all $M$ material areas to the surface. Here, $d_j(\mathbf{r})$ denotes the distance to the surface of an area $j$, depending on the point of excitation within the grating structure $\mathbf{r}$. The fluorescence intensity of emission line $l$ calculated this way is proportional to the count rate $F_l(\alpha_\mathrm{i})$ which can be measured. A silicon-drift detector (SDD) measures photons within its detection angle and sensitivity range (Fig.~\ref{fgr:hybrid_scheme}, right). From the raw data, the count rates $F_l(\alpha_\mathrm{i})$ of each fluorescence line $l$ are extracted through decon\-volution and background subtraction~\cite{Scholze:2006}. The total fluorescence intensity for emission line $l$ is then calculated as follows:
\begin{equation}
    \Phi_l(\alpha_\mathrm{i}) = \frac{4\pi\sin\alpha_\mathrm{i}}{\Omega(\alpha_\mathrm{i})}\frac{F_l(\alpha_\mathrm{i})}{N_0\epsilon(E_l)}\ ,
    \label{eq:fluorescence_intensity}
\end{equation}
where $\Omega(\alpha_\mathrm{i}) / 4\pi$ is the effective solid angle of detection, $N_0$ the incident photon flux and $\epsilon(E_l)$ the detection efficiency for the fluorescence photon energy $E_l$~\cite{Andrle:2023}.

\subsection{Instrumentation and Measurement}

\begin{figure}[ht]
 \centering
 \includegraphics[width=\columnwidth]{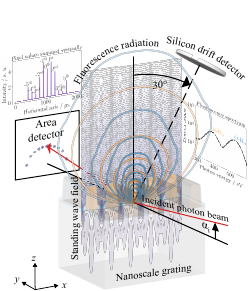}
 \caption{Scheme of the hybrid measurement approach of soft X-ray fluorescence scattero\-metry (XFS). A monochro\-matized photon beam hits a nano\-scale grating of different materials under a grazing angle of incidence $\alpha_\mathrm{i}$. Incoming photons are either scattered or absorbed, exciting different grating materials. The standing wave field determines diffraction into diffraction orders (peaks at the area detector) and stimulated emission (blue and orange contour lines over the sample surface). A silicon drift detector, $30^\circ$ off the plane of incidence, can measure the spectrum of the radiation from the sample surface, including the fluorescence emission lines (right).}
 \label{fgr:hybrid_scheme}
\end{figure}

The hybrid measurement approach utilizes the fact that some of the incoming photons are scattered at the grating structure while others get absorbed and excite atoms of the grating material if the incident photon energy has an appropriate value. The choice of the incident photon energy $E_\mathrm{i}$ is primarily determined by the material composition of the grating sample. Here, $E_\mathrm{i}$ should be significantly higher than the transition energies of the K$_{\alpha}$-shells of nitrogen and oxygen at $392.4\,\mathrm{eV}$ and $524.9\,\mathrm{eV}$~\cite{Zschornack:2007}, to excite the nitrogen and oxygen atoms and to be far from the next absorption edge. Thus, the incident photon energy is set to $680.0\,\mathrm{eV}$. By scanning over the critical angle of total external reflection, a strong scattering signal is replaced by a strong fluorescence signal through increasing absorption and stimulated emission. Figure~\ref{fgr:hybrid_scheme} shows a sketch of the hybrid measurement scheme of soft X-ray fluorescence scattero\-metry (XFS) for a fixed angle of incidence. The standing wave field determines the excitation within the grating structure. The stimulated emission is influenced by the grating structure through self-absorption, depending on the exit angle of the fluorescence photons (contour lines in Fig.~\ref{fgr:hybrid_scheme}).

\begin{figure}[ht]
 \centering
 \includegraphics[width=\columnwidth]{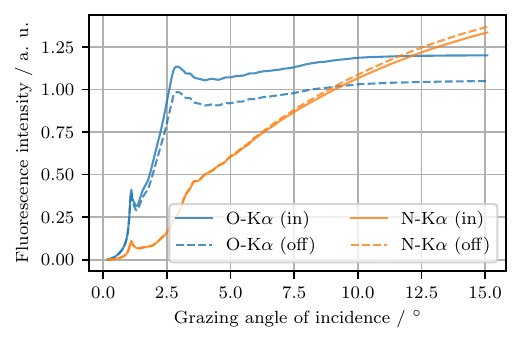}
 \caption{Fluorescence intensity calculated with self-absorption in and off the plane of incidence for N-K$_\alpha$ and O-K$_\alpha$ fluorescence from an \ce{Si3N4} grating model like shown in Fig.~\ref{fgr:grating_profile}b.}
 \label{fgr:ac_fluo}
\end{figure}

The XFS setup used in this work can perform simultaneous soft X-ray scattering and soft X-ray fluorescence measurements. A special feature of this instrument is that to change the grazing angle of incidence, the chamber rotates together with its permanently mounted area detector and SDD. The CCD camera mounted behind the sample detects elastically scattered soft X-rays while the SDD, $30^\circ$ off the plane of incidence, detects soft X-ray fluorescence (Fig.~\ref{fgr:hybrid_scheme}). Fluorescence radiation that leaves the grating surface under an angle of $30^\circ$ off the plane of incidence is influenced differently by self-absorption through the shadowing effect than that leaving the surface in the plane of incidence. The measurement of fluorescence radiation off the plane of incidence can be more sensitive to the shape of the grating than a conventional measurement within the plane of incidence. Thus, to increase the sensitivity of soft X-ray fluorescence, the work makes use of this effect. Figure~\ref{fgr:ac_fluo} shows the influence of the shadowing effect to the fluorescence intensity calculated for the N-K$_\alpha$ and O-K$_\alpha$ emission lines by comparing fluorescence intensity calculated in and off the plane of incidence. The comparison shows that the relative deviations for O-K$_\alpha$ fluorescence is about ten times lager than that for N-K$_\alpha$. This comes from the fact that the O-K$_\alpha$ fluorescence photons from the grating grooves need to go through the neighbor lines, while N-K$_\alpha$ fluorescence photons mainly leave the grating lines where they appear. Moreover, in cases where the lines of the grating are relatively high, like in this case, more fluorescence photons from inner regions of the lines can be detected if the pitch is sufficiently large. Figure~\ref{fgr:ac_fluo} shows that up to around $1\%$ more N-K$_\alpha$ fluorescence photons from the line can be detected.

The XFS setup can perform scans over the grazing angle of incidence $(0^\circ\dots 32^\circ)$. The azimuthal tilt angle of the sample can be changed for sample alignment. Limitations on the scan are given by the footprint of the photon beam on the sample surface as well as the measurement time given. The beam footprint on the sample surface can be kept small by using pinholes with small diameter or large incident angles. The distance between the $80\,\mathrm{\mu m}$-pinhole and the sample in the center of the chamber is $43\,\mathrm{cm}$. Thus, the photon beam cross-section with a diameter of around $83\,\mathrm{\mu m}$ at the sample and the resulting beam divergence of about $(0.2\times 0.2)\,\mathrm{mrad}^2$ (width by height) are primarily determined by Fraunhofer diffraction at the pinhole. The grazing angle of incidence determines the additional elongation of the photon beam footprint on the sample. For the angular scan, the lower limit of the grazing angle of incidence $(\alpha_\mathrm{i}\approx 2^\circ)$ is given by the over-illumination of the CCD due to being under the critical angle of total external reflection. At this angle, the beam footprint is still within the elongation of the grating lines at about $(0.08\times 2.4)\,\mathrm{mm}^2$ (width by elongation). The upper limit of the angle of incidence is determined by the lower limit of the signal-to-noise ratio of the higher-order diffraction efficiency. Thus, in this work, the scan area is reduced to the scan of the grazing angle of incidence from $2^\circ$ to $6^\circ$ under perfect conical sample mounting.

\begin{figure}[ht]
\centering
  \includegraphics[width=\columnwidth]{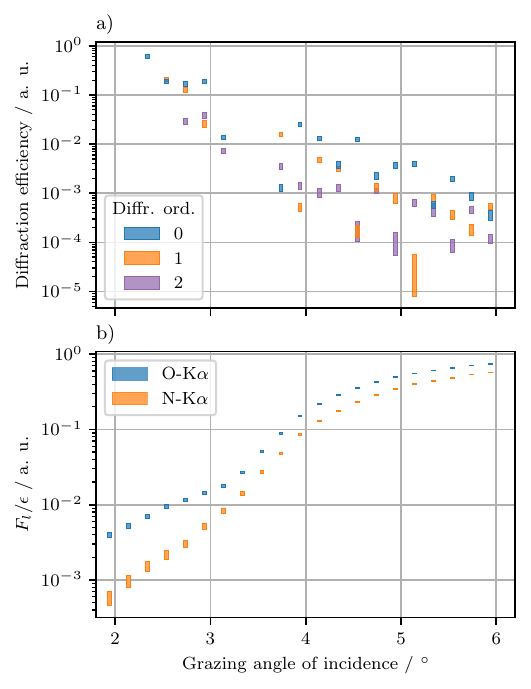}
  \caption{Measurement data simultaneously obtained as described in Fig.~\ref{fgr:hybrid_scheme} by a) soft X-ray scattering and b) soft X-ray fluorescence from the silicon nitride (\ce{Si3N4}) nano\-grating described in Fig.~\ref{fgr:grating_profile} over the grazing angle of incidence at an incident photon energy of $680.0\,\mathrm{eV}$. The uncertainty shown in a) and b) is $\pm 3\sigma$.}
  \label{fgr:scan_data}
\end{figure}

The CCD captures the diffraction pattern from the grating sample and yields the diffraction efficiency $I_m(\alpha_\mathrm{i}, E_\mathrm{i})$ for all diffraction orders $m$. To get a high signal-to-noise ratio staying in the regime of linearity between integrated photon counts and actual diffraction efficiency, for each diffraction efficiency several images with short exposure times are taken to extract the integrated diffraction counts from the regions around the diffraction peaks. Here, the normalized integrated photon counts can be identified with the diffraction efficiency in arbitrary units because the area detector is not calibrated. Figure~\ref{fgr:scan_data}a shows the diffraction efficiency for some orders of diffraction $m$ over the grazing angle of incidence. The total uncertainty of the normalized diffraction efficiency consists of the standard deviation of the signal over the combined diffraction images and the relative uncertainty of $2\%$ from detector inhomogeneity. As the measured diffraction efficiency has arbitrary units, the dimensional reconstruction of the nano\-scale grating uses the relative diffraction efficiency $\widetilde{I}_{m}$, which is the diffraction signal of higher order of diffraction $(m\neq 0)$ divided by the diffraction signal of the zeroth order of diffraction $(m=0)$ for each grazing angle of incidence. While collecting fluorescence spectra, the CCD is protected against over-illumination by a shutter in front of it.

Complementing the hybrid data set over the grazing angle of incidence, Fig.~\ref{fgr:scan_data}b shows the measured count rates over the detector efficiency $F_l / \epsilon(E_l)$ for the fluorescence lines of oxygen K$_\alpha$ and nitrogen K$_\alpha$.
Because of the reduced photon flux behind the $80\,\mathrm{\mu m}$-pinhole, the yield of fluorescence photons from the excited sample surface is relatively low. Thus, the integration time of the SDD is set to $1500$ seconds and the SDD is brought to less then $1\,\mathrm{mm}$ close to the sample surface to achieve a sufficient signal-to-noise ratio in the fluorescence spectra. Due to the small distance between detector and sample surface, fluorescence radiation from the sample surface over a large solid angle is detected. This explains the difference between the fluorescence curves of Fig.~\ref{fgr:ac_fluo} and Fig.~\ref{fgr:scan_data}b. The influence of the detector solid angle to the shape of the fluorescence intensity curve is described by the factor $4\pi\sin\alpha_\mathrm{i} / \Omega (\alpha_\mathrm{i})$ in Eq.~\ref{eq:fluorescence_intensity}. Here, the actual solid angle is unknown as the sample-detector distance cannot be accurately determined in the XFS setup. Thus, this work uses the relative fluorescence intensity $\widetilde{\Phi} = \Phi_{\mathrm{N-K}_\alpha}/\Phi_{\mathrm{O-K}_\alpha}$ for the dimensional reconstruction of the nano\-scale grating, where the factor in Eq.~\ref{eq:fluorescence_intensity} cancels out. The relative uncertainty of the count rate for both emission lines is determined by photon fluctuations $\sqrt{F_l} / F_l$ each.

Comparing scattering and fluorescence in the hybrid data set in Fig.~\ref{fgr:scan_data}, the measurement curves show that while the diffraction efficiency (Fig.~\ref{fgr:scan_data}a) decreases over $\alpha_\mathrm{i}$, the fluorescence intensities (Fig.~\ref{fgr:scan_data}b) from oxygen and nitrogen K$_\alpha$ increase. This general trend can be explained by the modes of the standing wave field which start to penetrate deeper and deeper into the grating over the grazing angle of incidence, with scattering decreasing and absorption increasing.

\section{Method}

To determine solutions for the sample-based parameters $\mathbf{P}_\mathrm{sample} = (h, w, \beta, r_\mathrm{bottom}, r_\mathrm{top}, d_\mathrm{groove}, d_\mathrm{line}, \xi)$ describing the shape of the nano\-scale grating's profile shown in Fig.~\ref{fgr:grating_profile} and the distribution of line edge and line width roughness, the inverse problem of describing the measurement data shown in Fig.~\ref{fgr:scan_data} can be solved using a model of the grating (Fig.~\ref{fgr:grating_profile}b) and of the experiment. Based on a set of parameters, the diffraction efficiency and fluorescence intensity can be calculated according to Eq~\ref{eq:diffraction_efficiency} and Eq.~\ref{eq:sherman} from the calculated electric field strength of the standing wave field. This work utilizes the computer software \textit{JCMsuite} (version 6.0.10) from the company JCMwave  GmbH, Berlin~\cite{JCMsuite,Burger:2012} to calculate the electric field strength of the standing wave field as well as the fast Fourier transform for the diffraction efficiency and the density integration for the fluorescence intensity. In a global optimization process, measured and calculated relative fluorescence intensities, $\widetilde{\Phi}_i^\mathrm{meas}$ and $\widetilde{\Phi}_i^\mathrm{calc}$, and diffraction efficiency, $\widetilde{I}_{m,i}^\mathrm{meas}$ and $\widetilde{I}_{m,i}^\mathrm{calc}$, can be compared for varied parameters to solve the inverse problem.

An optimal hybrid dimensional reconstruction of the nano\-scale grating has the right balance of information from the data sets of soft X-ray scattering and soft X-ray fluorescence. A weighting parameter, denoted as $\gamma$, can balance an optimization by weighting a combined $\chi^2$-function to be minimized, denoted as $\chi_\gamma^2$-function. Assuming the measurement uncertainties are normal distributed and non-correlated, and soft X-ray scattering and soft X-ray fluorescence measurements are independent, the combined $\chi_\gamma^2$-function can be written as
\onecolumngrid
\begin{equation}
    \chi_\gamma^2 = \frac{\gamma}{\hat{\chi}_\mathrm{scat}^2}\sum_{i,m}\frac{\left(\widetilde{I}_{m,i}^\mathrm{meas} - \widetilde{I}_{m,i}^\mathrm{calc}\right)^2}{\sigma_{m,i}^2} + \frac{(1-\gamma)}{\hat{\chi}_\mathrm{fluo}^2}\sum_i\frac{\left(\widetilde{\Phi}_i^\mathrm{meas} - \widetilde{\Phi}_i^\mathrm{calc}\right)^2}{\sigma_{i}^2}\ ,
    \label{eq:chi2_gamma}
\end{equation}
\twocolumngrid
with the standard deviations of the Gaussian uncertainties for scattering, $\sigma_{m,i}$, and fluorescence, $\sigma_{i}$, and the final values of the $\chi^2$-functions $\hat{\chi}_\mathrm{scat}^2 = \sum_{i,m}(\widetilde{I}_{m,i}^\mathrm{meas} - \widetilde{I}_{m,i}^\mathrm{calc})^2/\sigma_{m,i}^2$ and $\hat{\chi}_\mathrm{fluo}^2 = \sum_i(\widetilde{\Phi}_i^\mathrm{meas} - \widetilde{\Phi}_i^\mathrm{calc})^2/\sigma_{i}^2$ for completed optimizations with soft X-ray scattering only and soft X-ray fluorescence only, respectively. The found $\chi^2$-function values, $\hat{\chi}_\mathrm{scat}^2$ and $\hat{\chi}_\mathrm{fluo}^2$, represent the individual best-fits. The squared standard deviations are made up of the sum of the squared uncertainties of the measurement and the model. By normalizing the sums of the residuals for soft X-ray scattering and soft X-ray fluorescence with $\hat{\chi}_\mathrm{scat}^2$ and $\hat{\chi}_\mathrm{fluo}^2$, respectively, the $\chi_\gamma^2$-function compensates the effect of different sizes, scales and uncertainties between the individual data sets to the weighting parameter $\gamma$. To find the right balance for the hybrid dimensional reconstruction, optimization results from a series of different parameter values $\gamma\in (0,1)$ can be compared by their final $\chi_\gamma^2$-function values.

In this work, the numerical precision of the model of the standing wave field is limited to reduce computational effort. The limited precision can have a sensitive effect to the model uncertainty of the diffraction efficiency $I_m(\alpha_\mathrm{i})$ for some grating shapes and grazing angles of incidence. As the calculated far-field is the fast Fourier transform of the whole computational domain, this can lead to an overall model uncertainty of up to $15\%$. The calculated fluorescence intensity $\Phi_{l,j}(\alpha_\mathrm{i})$, on the other hand, results from an averaging of excitation over parts of the domain. Thus, its model uncertainty is smaller than that of the diffraction efficiency, limited by $3\%$. The optical constants for \ce{Si3N4} and \ce{SiO2} are taken from a reconstruction of a layer system by means of soft X-ray reflectivity~\cite{Andrle:2021}, while those of \ce{Si} are taken from data base~\cite{Henke:1993}. Values for the partial photoioni\-zation cross-section $\tau (680\,\mathrm{eV})$ and the fluorescence yield $\omega_k$ of the nitrogen K$_\alpha$-shell and the oxygen K$_\alpha$-shell, are taken from source~\cite{Elam:2002}, respectively. To account for small angular displacements of the grazing angle of incidence $\alpha_\mathrm{i}$ and the azimuthal tilt angle of the sample $\varphi$ in the dimensional reconstruction, setup-based parameters $\mathbf{P}_\mathrm{setup} = (\Delta\alpha_\mathrm{i},\Delta\varphi)$ can determine the actual angular orientation of the grating with respect to the incoming photon beam by $\alpha_\mathrm{i}\prime = \alpha_\mathrm{i} + \Delta\alpha_\mathrm{i}$ and $\varphi\prime = \varphi + \Delta\varphi$. This work uses differential evolution~\cite{Storn:1997,SciPy:2020} to minimize the $\chi_\gamma^2$-function, described in Eq.~\ref{eq:chi2_gamma}, with constant mutation rate of $80\%$ and recombination rate of $50\%$. Parameter values are sampled by individual optimizations over $161$ different weighting parameter values $\gamma$ between $0$ (fluorescence only) and $1$ (scattering only) using the same parameter ranges and initial values. The solutions found can accumulate in clusters, which can be analyzed by principal component analysis (PCA)~\cite{Pearson:1901,Hotelling:1933} and k-means clustering~\cite{MacQueen:1967,Lloyd:1982,Forgy:1965}, using the implementation for PCA and the KMeans class provided by the \textit{Scikit-learn} library (version 1.2.2)~\cite{Pedregosa:2011}.

\section{Results and discussion}

\begin{figure}[ht]
\centering
  \includegraphics[width=\columnwidth]{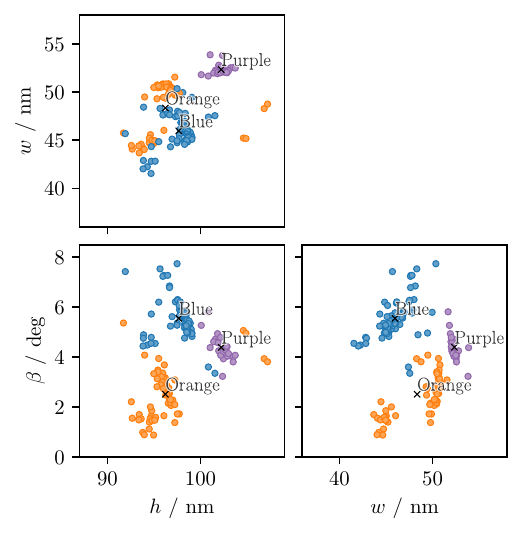}
  \caption{Solutions of the  dimensional reconstruction of the \ce{Si3N4} nano\-grating shown in Fig.~\ref{fgr:grating_profile} found by minimizing Eq.~\ref{eq:chi2_gamma} for $161$ weighting parameter values between $\gamma = 0$ (fluorescence only) and $\gamma = 1$ (scattering only). The solutions accumulate in clusters marked with colors.}
  \label{fgr:scatter_corner_plot}
\end{figure}

When optimizing over the weighting parameter $\gamma$, the values of the grating parameters $(\mathbf{P}_\mathrm{sample})$ vary over the prior ranges, listed in Tab.~\ref{tbl:cluster_metrics}. The variation of these values can be explained by the two facts that, first, differential evolution might not find the global minimum in a possibly multi\-modal solution space and, second, might stop close to a minimum due to the defined termination criterion. Three clusters can be identified (Cluster Orange, Cluster Blue and Cluster Purple) using PCA with $3$ components and k-means clustering with a fixed number of clusters, here $3$. Figure~\ref{fgr:scatter_corner_plot} shows the best-fit solutions of the most relevant grating parameters labeled with different colors (orange, blue, purple) according to the clusters to which they belong. The clusters cover different areas in which multi\-modal best-fit solutions lie. Cluster Orange and Cluster Blue have relatively large spreads compared to that of Cluster Purple. Of Cluster Orange and Cluster Blue, Cluster Orange has the largest spread. Comparing characteristics like the spread of best-fit solutions can indicate how sensitive parameter values within a cluster depend on the weighting in the $\chi_\gamma^2$-function. A relative weak dependency of parameter values in a cluster can indicate whether the actual solution of the geometry of the nano\-scale grating lies in this cluster.

\begin{table*}
\small
  \caption{\ Mean values and relative ranges of the \ce{Si3N4} nano\-grating parameters in the solution clusters (Fig.~\ref{fgr:scatter_corner_plot})}
  \label{tbl:cluster_metrics}
  \begin{tabular*}{\textwidth}{@{\extracolsep{\fill}}cccccccccc}
    \hline
    \multicolumn{4}{r}{} & \multicolumn{2}{c}{Cluster Orange} & \multicolumn{2}{c}{Cluster Blue} & \multicolumn{2}{c}{Cluster Purple}\\
    \hline
    Parameter & \multicolumn{3}{c}{Prior range} & Mean & Rel. range / \% & Mean & Rel. range / \% & Mean & Rel. range / \% \\
    \hline
    &&&&&&&&&\\
    $h$ / nm & $87$ & \dots & $109$ &                96.2 & 16 & 97.7 & 10 & 102.2 & 4 \\
    $w$ / nm & $36$ & \dots & $58$ &                 48.3 & 16 & 46.0 & 19 & 52.3 & 4 \\
    $\beta$ / $^\circ$ & $0$ & \dots & $8.5$ &            2.5 & 179 & 5.6 & 79 & 4.4 & 59 \\
    $r_\mathrm{bottom}$ / nm & $1$ & \dots & $21$ &  3.5 & 336 & 17.4 & 107 & 17.5 & 19 \\
    $r_\mathrm{top}$ / nm & $1$ & \dots & $15$ &     10.4 & 59 & 7.3 & 76 & 11.3 & 32 \\
    $d_\mathrm{g}$ / nm & $1.32$ & \dots & $14.82$ & 9.1 & 62 & 6.3 & 115 & 11.2 & 30 \\
    $d_\mathrm{l}$ / nm & $1.34$ & \dots & $4.84$ &  3.6 & 21 & 2.8 & 53 & 3.0 & 21 \\
    $\xi$ / nm & $0$ & \dots & $5$ &                 0.3 & 492 & 0.5 & 339 & 2.1 & 139 \\
    &&&&&&&&&\\
    \hline
    \multicolumn{4}{r}{Cluster range mean $\rightarrow$} & & 148 &  & 100 &  & 38\\
    \hline
  \end{tabular*}
\end{table*}

\begin{figure}[ht]
\centering
  \includegraphics[width=\columnwidth]{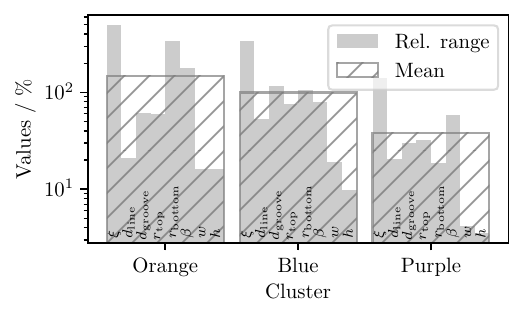}
  \caption{Comparison of the relative ranges of the \ce{Si3N4} nano\-grating parameters listed in Tab.~\ref{tbl:cluster_metrics} across the solution clusters shown in Fig.~\ref{fgr:scatter_corner_plot}.}
  \label{fgr:cluster_metrics}
\end{figure}

Figure~\ref{fgr:cluster_metrics} shows the ranges of the parameter values which are normalized to their mean values for each cluster. The mean values and relative parameter ranges are listed in Tab.~\ref{tbl:cluster_metrics} together with the prior ranges of the optimization. The mean of the relative ranges of all parameters can be calculated to compare the clusters in terms of the dependency of solutions on the weighting in the $\chi_\gamma^2$-function. The relative ranges and their mean values support the observed spread of the clusters from Fig.~\ref{fgr:scatter_corner_plot}. Cluster Purple has the smallest relative ranges which indicates that a solutions with a line height of about $h = 102.2\,\mathrm{nm}$ are least influenced by the weighting in the $\chi_\gamma^2$-function. This comparison indicates that the actual solution of the grating geometry is in the parameter ranges of Cluster Purple.

\begin{figure}[ht]
\centering
  \includegraphics[width=\columnwidth]{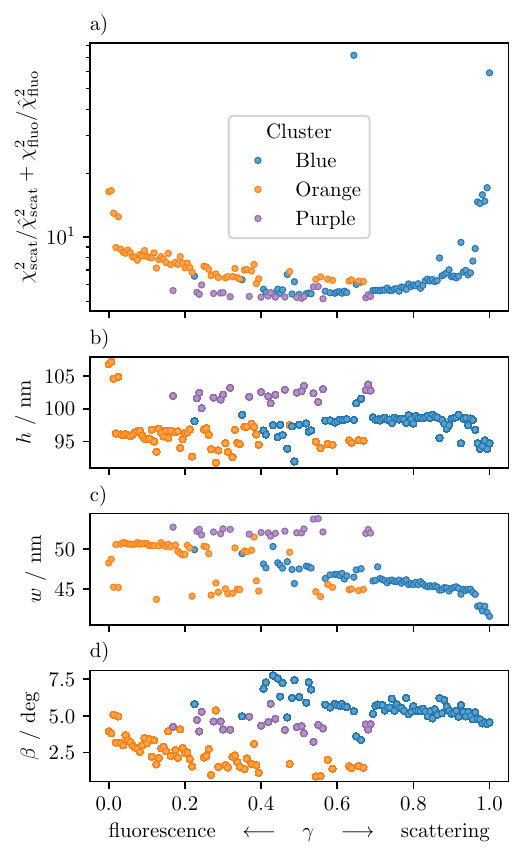}
  \caption{Values of the objective function $\chi_\gamma^2$ used for dimensional reconstruction of the nano\-scale grating shown in Fig.~\ref{fgr:grating_profile} over the weighting parameter $\gamma$ used in Eq.~\ref{eq:chi2_gamma}. The red dot indicates a local minimum that indicates an optimal value of $\gamma$ for hybrid dimensional reconstruction of the grating.}
  \label{fgr:chi2_gamma}
\end{figure}

To further analyze the clusters, cluster labels can be assigned to the $\chi^2$-function values of the combined regression. Figure~\ref{fgr:chi2_gamma}a shows the sum of the normalized $\chi^2$-function values of soft X-ray scattering only, $\chi_\mathrm{scat}^2$, and soft X-ray fluorescence only, $\chi_\mathrm{fluo}^2$, of the best-fit solutions sampled over the weighting parameter $\gamma$, marked according to the solution clusters $\left(\chi_\mathrm{scat}^2 / \hat{\chi}_\mathrm{scat}^2 + \chi_\mathrm{fluo}^2 / \hat{\chi}_\mathrm{fluo}^2\right)$. Solutions in Cluster Orange and Cluster Blue can be interpreted as solutions found under the dominance of one of the two data sets in the optimization. Around the central value of the weighting parameter $(\gamma = 0.5)$, solutions are influenced by the two parts of the hybrid data set, which results in better compromises for fitting the hybrid data set and the appearance of solutions that belong to Cluster Purple. This confirms that soft X-ray scattering and soft X-ray fluorescence data are complementary. The solutions of Cluster Purple appear excellent compared to those of Cluster Orange and Cluster Blue because they find the best compromise of fitting the hybrid data set, while also displaying an independence from the weighting of the fit. Figures~\ref{fgr:chi2_gamma}b-d show the direct dependency of the solutions represented by line height $h$, line width $w$ and sidewall-angle $\beta$ from the weighting parameter $\gamma$. While the parameter values from the Cluster Orange and Cluster Blue drift and jump strongly over $\gamma$, those values from Cluster Purple only fluctuate around the mean value.

An appropriate value for weighting the combined regression can be found in the center region from around $\gamma \approx 0.16$ to $0.67$, where the chance to find the actual solution, assigned to Cluster Purple, is relatively high. The fact that the center region is shifted off the center $(\gamma = 0.5)$ towards smaller $\gamma$-values could be explained by the sensitivity of the soft X-ray scattering measurement to the angular position which is in this case higher than that of soft X-ray fluorescence. To compensate the different sensitivities of the measurements or the relative information content of the data sets, the value of the weighting parameter should be set in such a way that the influence of the one data set with less relative information content than that of the other data set is higher. Otherwise the data set with higher relative information content would dominate the optimization process which probably gives the dominating multimo\-dality as solution.

\begin{figure*}[ht]
 \centering
 \includegraphics[width=\textwidth]{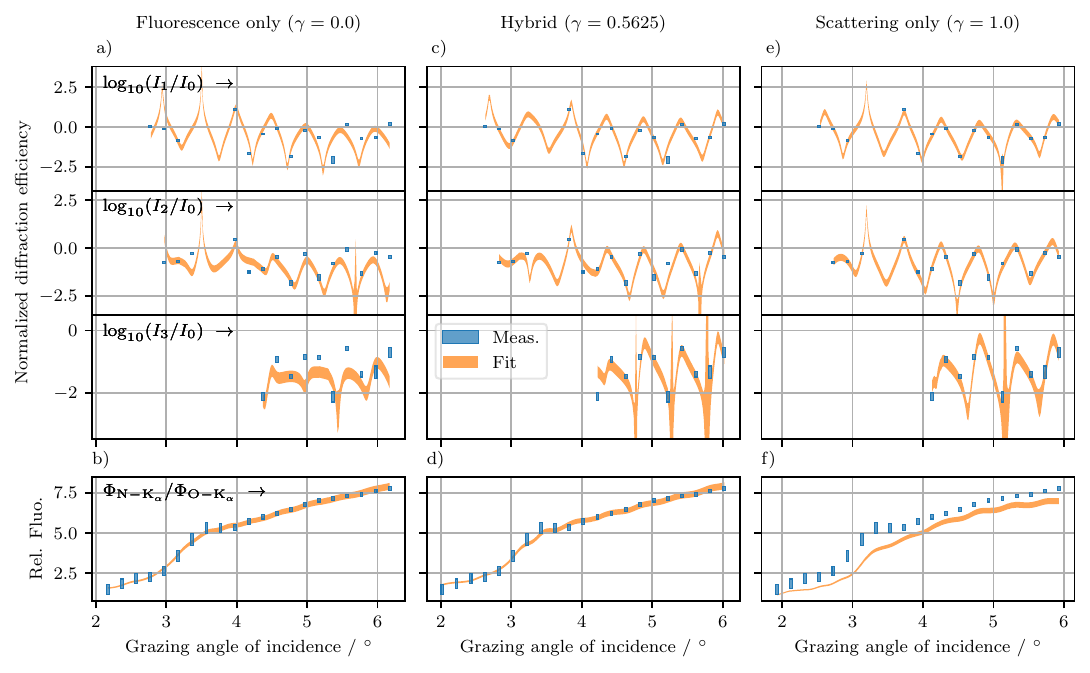}
 \caption{Optimization results for soft X-ray scattering and fluorescence with different values of the weighting parameter $\gamma$ from Eq.~\ref{eq:chi2_gamma}, separated by columns ($\gamma = 0$ fluorescence only, $\gamma = 0.5625$ hybrid and $\gamma = 1$ scattering only). The first three rows contain the results for soft X-ray scattering over the grazing angle of incidence at incident photon energy $E_\mathrm{i}=680\,\mathrm{eV}$, represented by the relative diffraction efficiency $\widetilde{I}_{m,i}(\alpha_\mathrm{i},E_\mathrm{i})$ from $1^\mathrm{st}$ row for $1^\mathrm{st}$ over $0^\mathrm{th}$ diffraction order to $3^\mathrm{rd}$ row for $3^\mathrm{rd}$ over $0^\mathrm{th}$ diffraction order). The relative diffraction efficiency is rescaled by the deca\-dic logarithm. In the last row, b), d) and f) contain the results for soft X-ray fluorescence, represented by the relative fluorescence $\widetilde{\Phi}$ from nitrogen K$_\alpha$ over oxygen K$_\alpha$ over the grazing angle of incidence at incident photon energy $E_\mathrm{i}=680\,\mathrm{eV}$.}
 \label{fgr:results}
\end{figure*}

For a comparison of the results, Fig.~\ref{fgr:results} shows optimization results based on a fit of fluorescence data only $(\gamma=0)$ in the first column (Fig.~\ref{fgr:results}a-b), a fit of hybrid data set in the second column using $\gamma_\mathrm{hybrid} = 0.5625$, assigned to Cluster Purple (Fig.~\ref{fgr:results}c-d) and a fit of scattering data only $(\gamma=1)$ in the third column (Fig.~\ref{fgr:results}e-f). Here, the optimization results based on soft X-ray scattering and soft X-ray fluorescence data have a solution that fits only to the scattering and the fluorescence data set, respectively, while the result based on the hybrid data set fits to both scattering and fluorescence data, although not as well as to the individual data sets. Possible reasons for that could be an insufficient best-fit or physical model like the influence of the roughness of the grating lines to soft X-ray fluorescence which is not taken into account here.

Other hybrid metrology approaches compare and discuss results from different techniques~\cite{Murataj:2023,Yan:2006}, make use of combined $\chi^2$-functions for combined regression~\cite{Hansen:2022,Melhem:2023,Siaudinyte:2023,Henn:2015} or use the results of one characterization technique as Bayesian input for another technique~\cite{Silver:2011,Silver:2014}. The hybrid metrology approach in this work uses the combined regression and also takes the relative complexity or information content of data sets of different techniques into account. One possible reason for different results from different characterization techniques applied on the same sample might be the fact that the relative information content in the datasets differ from each other. The relative information content may depend on the number of data points and the sensitivity of the technique in certain measurement sections. Weighting the combined $\chi^2$-function with a parameter, while excluding the effect of the sizes of data sets and the measurement uncertainties to the dimensional reconstruction, can optimize the dimensional reconstruction and show the distribution of information over different data sets for hybrid metrology.

\section{Conclusions}

This work uses a setup for soft X-ray fluorescence scattero\-metry, combining soft X-ray scattering and soft X-ray fluorescence in a hybrid approach for characterizing a nano\-scale grating made of different low-Z materials with increased unam\-biguity of the result. The dimensional reconstruction of the nano\-scale grating is based on complementary data sets and uses combined regression. To compensate for differences in the sensitivity of the measurements or to equalize the relative information content of the data sets, the combined $\chi^2$-function used in the optimization is weighted via a parameter. This work shows that the weighting parameter has a large influence on the determined line profile of the grating. This confirms that soft X-ray scattering and soft X-ray fluorescence are complementary techniques. Moreover, only particular solutions based on the combined data set seem to be largely independent of the weighting. The value of the weighting parameter can give an indication of the relative information content of the complementary data sets in relation to the model used. This work shows the importance of a weighting parameter for a combined regression if the information content of the complementary data sets is unknown and its applicability to find the actual solution of the grating line profile over ambiguous solutions.

\section*{Conflicts of interest}
There are no conflicts to declare.

\section*{Acknowledgments}
The project (grant agreement number 101096772) is supported by Chips Joint Undertaking and its members, including the top-up funding of Belgium and the Netherlands.

The authors thank Thomas Siefke and his colleagues from the University of Jena for cleaning the \ce{Si3N4} grating sample. The authors also thank Philipp Hönicke from the Helmholtz-Zen\-trum Berlin (HZB) for fruitful discussion.

\bibliographystyle{apsrev4-2}
\bibliography{sample}

\end{document}